\documentclass{article}
\usepackage{hiph-art}

\volnumber{00} \issuenumber{0} \edyear{2005}                             
\frompage{000} \topage{000}                                              
\recrevdate{November 2005}                                               
\def\rightmark{A Modified ``Bottom-Up'' Thermalization}
\def\leftmark{A.H. Mueller, A.I. Shoshi and S.M.H. Wong}

\def\be{\begin{equation}}
\def\ee{\end{equation}} 
\def\bea{\begin{eqnarray}}
\def\eea{\end{eqnarray}}

\def\del{\partial} 

\def\a{\alpha}
\def\d{\delta} 
\def\l{\lambda} 
\def\r{\rho} 
\def\s{\sigma} 
\def\t{\tau}

\title{A Modified ``Bottom-up'' Thermalization in Heavy Ion Collisions} 
\authors{ 
{A.H. Mueller$^{1,a}$, A.I. Shoshi$^{1,2}$ and S.M.H. Wong$^1$
\index{Mueller, A.H.} 
\index{Shoshi, A.I.} 
\index{Wong, S.M.H.} 
}\\[2.812mm]
{\normalsize
\hspace*{-8pt}$^1$
Department of Physics, Columbia University, New York, NY 10027, 
U.S.A.\\[0.2ex] 
\hspace*{-8pt}$^2$
Fakult\"at f\"ur Physik, Universit\"at Bielefeld, 33615 Bielefeld, Germany 
\\[0.2ex] 
}}
 
\abstract{In the initial stage of the bottom-up picture of 
thermalization in heavy ion collisions, the gluon distribution is
highly anisotropic which can give rise to plasma instability. 
This has not been taken account in the original paper. It is shown 
that in the presence of instability there are scaling solutions,
which depend on one parameter, that match smoothly onto the late 
stage of bottom-up when thermalization takes place. 
}


 
\makeindex 
\begin{document}
 
\maketitle

\section{The original bottom-up picture}
\label{bottom-pic}

In the McLerran-Venugopalan model of the color glass condensate \cite{mv},
small-$x$ gluons with transverse momentum below a certain saturation 
scale $Q_s$ are at their maximum density. When applied to a nucleus-nucleus 
collision at impact parameter $b$, this scale is given by \cite{muell} 
\be Q^2_s = \frac{8 \pi^2 \a N_c}{N^2_c-1} \, \sqrt{R^2_A-b^2} 
            \, \r\, x G_p(x,Q^2_s) 
\ee
and its value is $Q_s \sim 1$ GeV at the relativistic heavy ion collider
(RHIC). Here $R_A$ is the nuclear radius, $\r$ is the nuclear number
density, $N_c$ is the number of color, $\a$ is the coupling and 
$G_p$ is the gluon distribution of a proton. In a 
nuclear collision these gluons have a typical momentum of $Q_s$
and are freed at a time around $1/Q_s$ after the initial impact. 
In the bottom-up picture, which is based on the observation that 
inelastic processes are no less important than elastic processes 
for thermalization \cite{sw}, equilibration is driven by these hard 
gluons and it goes through three distinct stages \cite{bmss}.  
They are a) the early times $1 < Q_s \t < \a^{-3/2}$, b) the 
intermediate times $\a^{-3/2} < Q_s \t < \a^{-5/2}$ and c) the final 
stage $\a^{-5/2} < Q_s \t < \a^{-13/5}$.

\subsection{a) $1 < Q_s \t < \a^{-3/2}$} 

At the early times hard gluons dominate 
and because of the longitudinal expansion the density goes down like
\be N_h \sim \frac{Q^3_s}{\a (Q_s \t)} \,. 
\ee 
In the central collision region most of the gluons have small 
longitudinal momentum, $p_z \ll 1$, otherwise they would have wandered
out of the region. But this momentum cannot be zero either because 
of broadening due to multiple scattering. Effectively the $p_z$ 
goes through a random walk in momentum space due to the random kicks
by other hard gluons so
\be p^2_z \sim N_{\textrm{\scriptsize col}}\, m^2_D 
          \sim \frac{\a N_h}{p_z}  
\ee
where $N_\textrm{\scriptsize col}$ is the number of collisions a hard 
gluon typically has encountered at the time $\t$ and $m^2_D$ is the 
screening mass square
\be m^2_D \sim \a \int d^3 p \frac{f_h(p)}{p} \sim \frac{\a N_h}{Q_s} 
          \sim \frac{Q^2_s}{Q_s \t} \,. 
\label{eq:m_D} 
\ee
which effectively acts as the variance for each kick due to the 
much more frequent small angle collisions. $p_z$ comes out to be 
\be p_z \sim (\a N_h)^{1/3} \sim \frac{Q_s}{(Q_s \t)^{1/3}} \,. 
\ee
Soft gluons with momentum $k_s$ are produced during these times via 
the Bethe-Heitler formula \cite{gb} to give the parametric form for 
$N_s$ 
\be N_s \sim \t \frac{\del N_s}{\del \t} 
        \sim \frac{Q^3_s}{\a (Q_s \t)^{4/3}} \,. 
\ee
Once produced, random scattering by other gluons energizes these soft
gluons so that their momenta settle around $k_s \sim p_z$. Therefore 
the soft gluon distribution becomes 
\be f_s \sim \frac{N_s}{k_s^3} \sim \frac{1}{\a (Q_s \t)^{1/3}} \,.
\ee

\subsection{b) $\a^{-3/2} < Q_s \t < \a^{-5/2}$} 

In the intermediate times hard gluons still dominate in 
numbers but now $f_h < 1$. This changes the scattering rate 
with the hard gluons so 
\be k^2_s \sim N_{\textrm{\scriptsize col}}\, m^2_D 
          \sim \a\, Q^2_s  
\ee 
is now a constant. Assuming that the screening is mainly due to
the soft gluons 
\be m^2_D \sim \frac{\a N_s}{k_s} \gg \frac{\a N_h}{Q_s} \,,
\ee
one can find self-consistently that 
\be N_s \sim \frac{\a^{1/4} Q^3_s}{(Q_s \t)^{1/2}} \,. 
\ee

\subsection{c) $\a^{-5/2} < Q_s \t < \a^{-13/5}$} 
 
In the final stage most gluons are soft $N_s \gg N_h$. 
The remaining hard gluons will scatter with 
the soft gluons and lose energy via successive gluon splitting.
Whereas in the previous stages gluon production via the Bethe-Heitler 
formula is unaffected by multiple scattering, this is no longer true 
as the branching gluon momenta now fall within the range of the 
Landau-Pomeranchuk-Migdal suppression \cite{lpm}. Specifically 
gluon emission with momentum larger than 
$k_{\textrm{\scriptsize LPM}} = m^2_D/N_{\textrm{\scriptsize scatt}} \s$ 
is suppressed \cite{bsz}. $N_{\textrm{\scriptsize scatt}}$ is the number
density of the particles that is responsible for most of the scatterings. 
In this case the formation time of the branching gluon is 
$t_f \sim k_{\textrm{\scriptsize br}}/k_t^2$ where $k_t$ is the 
transverse momentum picked up by the branching gluon through the random
kicks by the soft gluons. It can be estimated as momentum broadening
as before but the number of collisions is now restricted by the 
formation time $t_f$ and the mean free path $\l$, hence
\be k^2_t \sim m^2_D\, t_f/\l \,. 
\ee
The rate of branching is roughly related to the formation time via 
$1/t_{\textrm{\scriptsize br}} \sim \a /t_f$. Equating 
$t_{\textrm{\scriptsize br}}$ with $\t$ and requiring that the
soft gluon now be in a thermal bath $N_s \sim T^3$, one finds the
branching momentum to be 
\be k_{\textrm{\scriptsize br}} \sim \a^4 T^3 \t^2 \,. 
\label{eq:k_br} 
\ee
Lastly equating the energy flow from the hard gluons to the soft
thermal bath, the temperature is determined to have the linear
time dependence
\be T \sim \a^3 Q^2_s \t \,.
\ee 
We will see later on that how some of these parametric dependences are 
recovered even after instability is included into the consideration.

\section{The instability}
\label{instab}  

As mentioned previously, early on in the collision only small-$x$ 
gluons can remain in the central region and they have typical 
transverse momentum of the order of $Q_s$. This describes a picture
of gluons with highly anisotropic initial momentum distribution. 
In such a situation as pointed out a long time ago \cite{mrow} 
and more recently within the context of the bottom-up picture 
\cite{alm}, it would give rise to plasma instability. The instability
occurs because the dispersion relation for the soft gluons gives a 
negative value for the screening mass square 
\be m^2_D \sim - \frac{\a N_h}{Q_s}  
\ee
when the momentum distribution is highly anisotropic \cite{masssq}. 
Modes with momentum $k < m_D$ are unstable. For recent reviews on
the topic of instability in the context of heavy ion collisions, 
one can read for example \cite{mrow2}. Although the growth is 
exponential in nature$^b$ and should be very fast on the time scale of
$\t \sim 1/m_D$ or $Q_s \t \sim 1$, it is difficult for it to lead 
directly to equilibration because first the instability only produces
soft particles and second Arnold and Lenaghan (the first paper of 
\cite{masssq}) showed that equilibration cannot occur before 
$Q_s \t \sim \a^{-7/2}$ which is much later than $Q_s \t \sim 1$
for small $\a_s$. 
 
Instability creates many soft gluons as a result. There are two
possibilities for the system to evolve further:
\begin{itemize}

\item[(i)]{When the soft particles are saturated at $f_s \sim 1/\a$ 
further production via the instability will result in gluons with
$k \sim m_D$ being transferred to higher momenta.}

\item[(ii)]{Or the instability will be completely eliminated by the
soft gluons at saturation.}

\end{itemize}
In either case, in the same spirit of the bottom-up picture, it is 
natural to look for a scaling solution which connections the end 
of the exponential growth due to the instability to final 
equilibration.

\section{A possible scaling solution}
\label{scaling_soln}

The solution(s) that we propose of course still has to 
start with the longitudinally expanding initial hard gluons 
\be N_h \sim \frac{Q^3_s}{\a (Q_s \t)} \,.
\ee
For gluons produced sometimes after the beginning but before $\t$,
$1/Q_s < \t_0 < \t$ these have \cite{msw} 
\bea & & 
     N_s(\t,\t_0) \sim \frac{Q^3_s}{\a (Q_s \t) (Q_s \t_0)^{1/3-\d}} 
     \;\;,\;\;\;\; 
     k_s(\t_0) \sim \frac{Q_s}{(Q_s \t_0)^{1/3-2\d/5}} \,,   \nonumber \\
     & & \hspace{3.0cm} 
     \a f_s (\t,\t_0) \sim 
                  \frac{(Q_s \t_0)^{1/3+\d/5}}{(Q_s \t)^{2/3+2\d/5}}
\label{eq:soln_t0} 
\eea 
where $N_s(\t,\t_0)$ is the number density of particle produced at 
time $\t_0$ but measured at $\t$. 

For gluons produced at time $\t$, one can write down a family of
$\d$-parameter dependent scaling solutions \cite{msw}   
\bea & &
     N_s \sim \frac{Q^3_s}{\a (Q_s \t)^{4/3-\d}} 
     \;\;,\;\;\;\; 
     k_s \sim \frac{Q_s}{(Q_s \t)^{1/3-2\d/5}} \,,   \nonumber \\
     & & 
     \a f_s \sim \frac{1}{(Q_s \t)^{1/3+\d/5}} 
     \;\;,\;\;\;  
     m_D \sim \frac{Q_s}{(Q_s \t)^{1/2-3\d/10}} \,,
\label{eq:scal_soln}
\eea 
where $\d \ge 0$. At $\d=0$ they coincide with the initial parametric 
form of the original bottom-up picture described in the first section.  
The solutions obey 
\bea m^2_D & \sim & \frac{\a N_s}{k_s}                    \\ 
     N_s   & \sim & \t \frac{\a^3}{m^2_D} (N_s f_s)^2 
\label{eq:bh}                                             \\ 
     k^2_s & \sim & m^2_D \frac{\t}{\t_{\textrm{\scriptsize col}}}  
     \mbox{\hspace{1cm} with} \mbox{\hspace{1cm}} 
     \frac{1}{\t_{\textrm{\scriptsize col}}} 
     \sim \frac{\a^2}{m^2_D} N_s f_s 
\label{eq:ks}                                             \\ 
     \frac{1}{\t} & \sim & \frac{\a^2}{k^2_s} N_s f_s \,. 
\eea
Here $m_D$ at $\t$ is determined by soft gluons produced 
via the Bethe-Heitler formula in Eq.~(\ref{eq:bh}). Multiple 
scattering ensures that these gluons gain momentum until they reach 
a value around $k_s$ given by Eq.~(\ref{eq:ks}). Once there they
scatter once on the average so they are borderline as far
as reaching equilibrium.

\section{The value of $\delta$ and $m^2_D > 0$?}
\label{maths}
 
So far we have always given the mass $m_D$ a subscript of $D$ which
stands for the Debye screening mass but in all reality, we are 
uncertain about the sign of the mass square. In section \ref{instab} 
we pointed out that the initial momentum distribution was highly
anisotropic, thus some soft gluon modes were unstable. Looking 
at the problem only parametrically as done in the bottom-up
picture and also here would not help us ascertain the sign of $m^2_D$. 
More dynamical inputs are necessary. One can compare
the momentum distribution and from the degree of anisotropy deduce
whether $m^2_D$ is negative. But the problem is more complicated
than that. For example from Eq.~(\ref{eq:soln_t0}) the contribution 
of the gluons produced at $\t_0$ to the screening mass square is 
\be m^2_D(\t,\t_0) \sim \frac{\a N_s(\t,\t_0)}{k_s(\t_0)} 
                   \sim \frac{Q^2_s (Q_s \t_0)^{3\d/5}}{Q_s \t} \,. 
\label{eq:m2_t0}
\ee 
If $\t_0 \ll \t$ then this contribution is clearly negative because
$k_s(\t_0)$ is so dissimilar to $k_s(\t)$. However the contribution
is small compared to $m^2_D$, which as seen in Eq.~(\ref{eq:scal_soln}),
has the same expression as Eq.~(\ref{eq:m2_t0}) except $\t_0$ is $\t$ 
in this case. On the other hand if $\t_0 \sim \t$ then the gluon's 
momentum distribution tends to be isotropic and it is unlikely that
$m^2_D$ is negative. The more difficult case is $\t_0 < \t$ when
one can no longer be certain when the sizes of the contributions to
$m^2_D$ are comparable. It is here that the parameter $\d$ plays
a role since the ratio of the late gluon to the early gluon  
contribution goes like $(\t/\t_0)^{3\d/5}$. Larger value of $\d$
put more weight on the late-time gluons' contribution. Better 
considerations and calculations are necessary to determine the value
of $\d$.$^c$

\section{Matching onto bottom-up}
\label{match}

\begin{figure}[ht] 
\vspace*{1.1cm} 
\insertplot{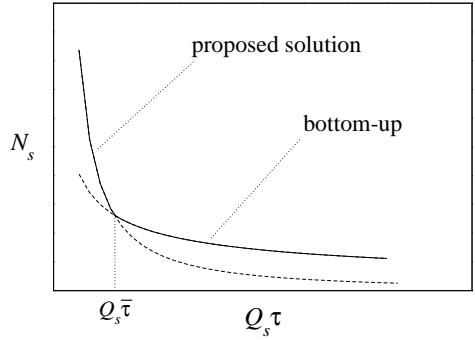} 
\vspace*{-2.5cm}
\caption{Graphical representation of the matching of the scaling 
solution to the bottom-up picture at $Q_s \bar \t$.}
\label{fig1}
\end{figure}

The solution(s) that we proposed in Eq.~(\ref{eq:scal_soln}) would not
be of any value if it did not describe also the equilibrium phase. 
In fact at a time when 
\be Q_s \bar \t \sim \a^{-15/2(5-6 \d)} 
\ee
our scaling solution becomes identical to the intermediate stage,
$\a^{-3/2} < Q_s \t < \a^{-5/2}$, of the bottom-up picture when  
the basic quantities in both cases go like 
\bea & &
     N_s \sim Q^3_s\, \a^{\frac{10-3\d}{2(5-6\d)}} 
     \;\;,\;\;\;\; 
     k_s \sim Q_s\, \a^{1/2} \,,                   \nonumber \\
     & & 
     f_s \sim \a^{\frac{5(-1+3\d)}{2(5-6\d)}} 
     \;\;\;\;,\;\;\;\;\;  
     m_D \sim Q_s\, \a^{\frac{3(5-3\d)}{4(5-6\d)}} \,.   
\label{eq:soln=BU} 
\eea 
This is true provided $0 < \d < 1/3$. A graphical representation
of this is shown in Fig. \ref{fig1}. 
At this time the present solution should make a transition
into the original bottom-up solution which remains true for
the rest of the evolution as long as the intermediate stage of the
bottom-up picture is not too affected by the initial presence of
the instability. 

For the case when $\d > 1/3$, we can see from Eq.~(\ref{eq:soln=BU})
that $f_s$ approaching unity. In fact in that case at a time 
$Q_s \t_1 \sim \a^{-15/(5+3\d)}$ already $f_s \sim 1$. Much of the 
picture of the final stage of the bottom-up becomes true except 
that gluons produced early at time $\t_0$ now play the part of the 
hard particles since $N_s(\t,\t_0) > N_h$ and $k_s(\t_0)$ now
functions as the branching momentum $k_{\textrm{\scriptsize br}}$ 
in Eq. (\ref{eq:k_br})  
\be k_s(\t_0) \sim \a^4 T^3 \t^2 \,.  
\label{eq:ks_br}
\ee 
The transfer of energy is similarly via gluon branching from these 
gluons into the bath of soft gluons. Equating once again the energy 
flow from these gluons into the thermal bath with temperature $T$ 
\be \frac{d \epsilon}{d \t} \sim T^3\, \frac{dT}{d\t} 
    \sim \frac{N_s (\t,\t_0)}{\t} \, k_s(\t_0) \,, 
\label{eq:dT/dt}
\ee 
using Eq.~(\ref{eq:soln_t0}) and Eq.~(\ref{eq:ks_br}) in 
Eq.~(\ref{eq:dT/dt}) one finds 
\be T \sim Q_s\, \a^{\frac{35-78\d}{39\d-10}} \, 
           (Q_s \t)^{\frac{15-36\d}{39\d-10}} \,.  
\label{eq:T_f} 
\ee
At $\d = 1/3$, this takes the familiar form $T \sim \a^2 Q^2_s \t$ 
of a linear increase of $T$ with $\t$ which is characteristic
of the bottom-up picture in \cite{bmss}. This heating up of the 
bath of soft gluons ends when the transfer of energy to the thermal 
bath is complete. This occurs when the branching momentum $k_s(\t_0)$
in Eq.~(\ref{eq:ks_br}) finally reaching $Q_s$ and 
\be T^4 \sim N_h(\t) \,.
\ee
At this time $Q_s \t \sim \a^{-13/5}$. Substituting this into 
Eq.~(\ref{eq:T_f}) one gets
\be T \sim Q_s \a^{2/5} \,,
\ee  
a value that is independent of $\d$. One sees that independent 
of what value $\d$ takes, as long as $\d > 1/3$, the scaling solutions  
match up to the final stage of bottom-up only at the final time 
$Q_s \t \sim \a^{-13/5}$.

\section*{Acknowledgment(s)}

A.S. acknowledges financial support by the Deutsche 
Forschungsgemeinschaft under contract Sh 92/2-1.

\section*{Notes} 
\begin{notes}
\item[a] The speaker. 

\item[b] In \cite{non-exp-beh} it was shown that at late times the
growth changed character from an exponential to a linear one and in
\cite{rv} for an longitudinally expanding plasma, the exponent
was shown to be $\sim \sqrt{\t}$ as one would expect from 
the form of Eq.~(\ref{eq:m_D}). 

\item[c] B\"odeker considered the broadening of the $p_z$ by 
multiple scattering with the much denser unstable gluon modes 
instead of with the hard gluons \cite{bod}. In that case he found
$p_z \sim Q_s/(Q_s \t)^{1/4}$ which would suggest a value for 
$\d \sim 5/24 < 1/3$ provided that $p_z$ takes this parametric 
form until the moment when the instability was finally eliminated. 
\end{notes}

\end{document}